**Small gap semiconducting organic charge-transfer interfaces**

M. Nakano,[1] H. Alves,[2] A. S. Molinari,[3] S. Ono,[1,4] N. Minder,[1] and A. F. Morpurgo[1*]

[1] *DPMC and GAP, University of Geneva, 24 quai Ernest-Ansermet, CH1211 Geneva, Switzerland*

[2] *INESC-MN and IN, Rua Alves Redol 9, 1000-029 Lisboa, Portugal*

[3] *Kavli Institute of Nanoscience, Delft University of Technology, Lorentzweg 1, 2628CJ Delft, The Netherlands*

[4] *Central Research Institute of Electric Power Industry, Komae, Tokyo 201-8511, Japan*

We investigated transport properties of organic heterointerfaces formed by single-crystals of two organic donor-acceptor molecules, tetramethyltetraselenafulvalene (TMTSF) and 7,7,8,8-tetracyanoquinodimethane (TCNQ). Whereas the individual crystals have un-measurably high resistance, the interface exhibits a resistivity of few tens of MegaOhm with a temperature dependence characteristic of a small gap semiconductor. We analyze the transport properties based on a simple band-diagram that naturally accounts for our observations in terms of charge transfer between two crystals. Together with the recently discovered tetrathiafulvalene (TTF)-TCNQ interfaces, these results indicate that single-crystal organic heterostructures create new electronic systems with properties relevant to both fundamental and applied fields.

---

[*] Electronic mail: Alberto.Morpurgo@unige.ch

The electronic properties of interfaces between different organic molecular semiconductors are crucial for the operation of most devices in the field of plastic electronics. In organic light-emitting diodes, for example, drastic performance enhancement can be achieved by introducing multiple material layers to form interfaces that separately optimize the microscopic processes involved in the device operation, such as carrier injection and recombination.[1] Another example of a functional interface is provided by the so-called bulk heterojunction,[2] which is currently investigated to improve the efficiency of organic solar cells. In both cases, transport across the interface is the relevant process. However, by analogy with conventional inorganic semiconductors, it is expected that transport parallel to the organic-organic interface should also be of great interest. Indeed, inorganic heterostructures made of III-V or II-VI semiconductors hosting two-dimensional electron gases have been widely studied,[3,4] leading to impressive new physics (e.g., the discovery of the quantum Hall effect)[5] and applications (e.g., high-electron-mobility transistors).[6,7] Nevertheless, for organic semiconductors, this "lateral" type of heterostructures has remained vastly unexplored.

A recent example illustrating the occurrence of interesting new phenomena in organic heterostructures is provided by charge-transfer interfaces formed by laminating single crystals of organic donor-acceptor molecules, tetrathiafulvalene (TTF) and 7,7,8,8-tetracyanoquinodimethane (TCNQ).[8] Despite the fact that the individual crystals are insulating, their interface exhibits a high electrical conductance. The phenomenon originates from a large charge transfer from donor to acceptor, which causes the formation of an interfacial two-dimensional metallic conductor. Notably, the two-dimensionality makes a TTF-TCNQ interface behave differently from a bulk TTF-TCNQ complex, in which TTF and TCNQ molecules are arranged into de-coupled one-dimensional chains, causing the material to become a Peierls insulator at low temperature. This difference illustrates how interfaces can lead to phenomena that do not occur in the corresponding bulk materials.

To start broadening the scope of molecular materials used in organic charge-transfer interfaces, in this letter we report on the investigation of lateral transport at interfaces consisting of TCNQ and tetramethyltetraselenafulvalene (TMTSF) single-crystals. TMTSF is a well-known donor molecule, which has led to the discovery of the first organic superconductors.[9,10] In the TMTSF-TCNQ interfaces, we observe a conductance that is thermally activated with a small (~ 100 meV) activation energy. From the measured mobilities of charge carriers in the individual crystals in conjunction with the measured resistivity values, we estimate that the density of transferred charge is in the order of $10^{11}$ cm$^{-2}$ at room temperature, corresponding to less than 0.001 electrons per molecules, i.e. approximately three orders of magnitude smaller than what is found in the two bulk phases,[11] and decreasing in a thermally activated way with lowering temperature. We analyze these findings in terms of a simple band-diagram, and show that our observations are consistent with a picture based on non-interacting electrons which are thermally excited from the valence band of TMTSF into the conduction band of TCNQ.

Figure 1 (b) and (c) show the schematics of the device structure used in this study together with an optical microscope image of an actual device. The details of the fabrication are virtually identical to those described in Refs. 8 and 17. The interfaces are assembled on a polydimethylsiloxane (PDMS) flexible substrate by laminating a TMTSF and a TCNQ single crystal onto each other. Crystals of both molecules were grown from vapor phase as described previously,[8,17] and the surface mobilities of charge carriers were characterized by means of room-temperature field-effect transistor measurements [the hole mobility of TMTSF is ~ 2-4 cm$^2$/Vs and the electron mobility of TCNQ is ~ 0.5 cm$^2$/Vs (Refs. 8 and 17)]. In what follows, we describe the results of the temperature-dependent transport measurements, and compare them to similar measurements performed on TTF-TCNQ interfaces identical to those discussed in Ref. 8, to illustrate the different behavior.

Figure 2 (a) and (b) show typical current-voltage (*I-V*) curves of TMTSF-TCNQ and TTF-TCNQ interfaces in a two-terminal and a four-terminal configuration measured at room-temperature. The slight non-linearity originating from the contacts is visible in the *I-V* curve of a TMTSF-TCNQ interface measured in a two-terminal configuration, and is almost entirely suppressed in a four-terminal measurement. Figure 2 (c) shows the histogram of the four-terminal resistivity values of TMTSF-TCNQ interfaces, and compares them to the data obtained from TTF-TCNQ interfaces. For TMTSF-TCNQ interfaces, all the resistivity values are in the 10-100 M$\Omega$ range, corresponding to resistances much smaller than those of the individual crystals (that is tens of GigaOhms, or typically much larger). The one order of magnitude spread in values is most likely originating from the different quality of the interfaces, mainly due to the non-perfect control of the manual lamination process used for the interface assembly. A spread of similar magnitude is observed in TTF-TCNQ interfaces, where the resistivity ranges between 10 and 100 k$\Omega$. Since the mobilities of charge carriers in all the crystals used in this study (TCNQ, TTF and TMTSF) have comparable values ($\sim$ 1 cm$^2$/Vs), the large difference in the resistivity between TMTSF-TCNQ and TTF-TCNQ interfaces indicates that the density of charge carrier in TMTSF-TCNQ interfaces is approximately 1000 smaller than in the TTF-TCNQ case.

Additional useful information can be extracted by measuring the temperature dependence of the interface resistivity. Figure 3 (a) and (b) compare the evolution of the four-terminal *I-V* curves for TMTSF-TCNQ and TTF-TCNQ interfaces as a function of temperature. For a given voltage, the current in a TMTSF-TCNQ interface decreases with decreasing temperature from 300 K (red) to 200 K (blue), whereas for a TTF-TCNQ interface the *I-V* curves are almost temperature independent, exhibiting a small increase in the best samples as previously reported.[8] The temperature dependence of the resistivity for the two cases is summarized in the Arrhenius plot of Fig. 3 (c). The good linearity of the data for a TMTSF-TCNQ interface indicates that conduction at this interface is thermally activated [$\rho \propto \exp(E_a/k_BT)$]. Measurements on 7 different interfaces gave a value of activation energy $E_a$ ranging from 70 to 120 meV ($E_a = 120$ meV for the device shown in the figure).

As a first step to analyze the behavior of TMTSF-TCNQ interfaces, we consider a simple band-diagram. The alignment depicted in Figure 4 (a) is the one that we expect qualitatively for a TMTSF-TCNQ interface based on the results of the transport measurements. Far away from the interface, the Fermi energy ($E_F$) is located in the middle of the band-gap for both TMTSF and TCNQ, because these crystals are intrinsic semiconductors. Close to the interface, however, the electrostatic potential associated to the charge transferred from TMTSF to TCNQ causes the bands to bend, and $E_F$ is located in the middle of the highest occupied molecular orbital (HOMO) of TMTSF and the lowest unoccupied molecular orbital (LUMO) of TCNQ. In this picture, the activation energy $E_a$ observed in the transport experiments corresponds to half the difference between the HOMO level of TMTSF and the LUMO level of TCNQ. Indeed, the value of $E_a$ measured -approximately 100 meV- compares well with the energy difference of the molecular levels estimated from electrochemical measurements.[18] For comparison, Fig. 4 (b) shows the band-diagram in the case of a TTF-TCNQ interface, which is expected from the observed metallic nature of this interface.

To substantiate the interpretation based on the band-diagram shown in Fig. 4 (a), we next estimate the sheet charge density ($n_s$) accumulated at a TMTSF-TCNQ interface. For non-interacting particles, $n_s$ can be simply calculated by integrating the product of density of states and distribution function. For narrow band organic crystals, the density of states can be estimated as $N_s/w$, where $N_s$ ($\sim 5 \times 10^{14}$ cm$^{-2}$) is the surface density of the molecules and $w$ ($\sim$ 0.5 eV)[19,20] is the bandwidth of the corresponding band. Since $E_a$ is sufficiently larger than $k_BT$, the carrier statistics is described by the Boltzmann distribution, and the value of $n_s$ is:

$$n_s = \int_{E_a}^{\infty} (N_s/w)\exp(-E/k_BT)dE = N_s \frac{k_BT}{w}\exp(-E_a/k_BT).$$
(1)

With the measured typical value of $E_a = 100$ meV, $n_s$ is evaluated to be in the order of $10^{11}$ cm$^{-2}$ at room temperature. Using this value and the approximate mobility value of ~ 1 cm$^2$/Vs of TMTSF and TCNQ single crystals, we estimate a resistivity at room temperature of around 10 MΩ, which compares well with the lowest resistivity value that we measure experimentally [see Fig. 2 (c)]. Finding such a good agreement using a simple physical picture suggests that the proposed description in terms of a band-diagram correctly captures the essential aspects of charge transfer at TMTSF-TCNQ interfaces.

The estimated charge density at a TMTSF-TCNQ interface is more than three orders of magnitude lower than that of a bulk TMTSF-TCNQ complex.[11] In the bulk complex, as a result of the large amount of charge transferred, the density of charge carriers is comparable to the density of molecules, and it is known that in this case electron interactions and strong correlations play an important role. At a TMTSF-TCNQ interface, on the contrary, the much lower density of transferred charge indicates that charge carriers are spatially well separated, and electron correlations should not be relevant. We believe that this is the reason why our description based on a simple band-diagram for non-interacting particles describes well the amount of charge transferred, and why the measured semiconducting energy gap is in fair agreement with the electrochemical data.[18]

In summary, we have shown that TMTSF-TCNQ interfaces provide a second example of organic charge-transfer interfaces. Contrary to TTF-TCNQ interfaces with a two-dimensional metallic system, TMTSF-TCNQ interfaces behave as a small gap semiconductor, in which thermal excitation is needed to transfer charge. The electronic phase of the interface appears to be different from that of a bulk TMTSF-TCNQ complex, providing more indications that an interface between organic crystals is useful to create novel electronic systems.

We acknowledge H. Xie and I. G. Lezama for experimental help and fruitful discussions. H.A., S.O. and A.F.M. also acknowledge financial support from FCT, Grant-in-Aid for Scientific Research (20740213) from MEXT, and MaNEP, respectively.

**Figure captions**

FIG. 1. (Color online) (a) Structure of the molecules used in this work: TTF and TMTSF act as a donor, and TCNQ as an acceptor. (b) Schematic representation of a device used to study transport at charge-transfer interfaces. The broken line represents the interfacial region where mobile charge carriers are present. (c) Optical microscope image of a device based on a TMTSF-TCNQ interface, including the scheme of the measurement configuration.

FIG. 2. (Color online) Panel (a) and (b) show the *I-V* curves of TMTSF-TCNQ and TTF-TCNQ interfaces measured at room temperature in a two-terminal and a four-terminal configuration. (c) Histogram of the resistivity values of TMTSF-TCNQ and TTF-TCNQ interfaces measured in a four-terminal configuration, on more than 20 interfaces for each system. In both cases, the spread in values is approximately one order of magnitude.

FIG. 3. (Color online) The four-terminal *I-V* curves of (a) TMTSF-TCNQ and (b) TTF-TCNQ interfaces measured at different temperatures ranging from 300 K (red) to 200 K (blue) in 20 K steps. (c) The Arrhenius plot of the resistivity for both systems. The resistivity of TMTSF-TCNQ is thermally activated with activation energy $E_a$ (ranging between 70 and 120 meV; $E_a$= 120 meV for the device shown here).

FIG. 4. (Color online) Schematic of the simplified band-diagrams of (a) TMTSF-TCNQ and (b) TTF-TCNQ interfaces. For a TMTSF-TCNQ interface, the Fermi level lies in the gap between the HOMO of TMTSF and the LUMO of TCNQ, and charge transfer from TMTSF to TCNQ is thermally activated. In a similar diagram for a TTF-TCNQ interface, the HOMO of TTF is higher in energy than the LUMO of TCNQ, and charge transfer occurs

spontaneously. In all materials, the Fermi level away from the interface lies in the middle of the HOMO-LUMO gap, as it should be, since the molecular crystals are intrinsic semiconductors.

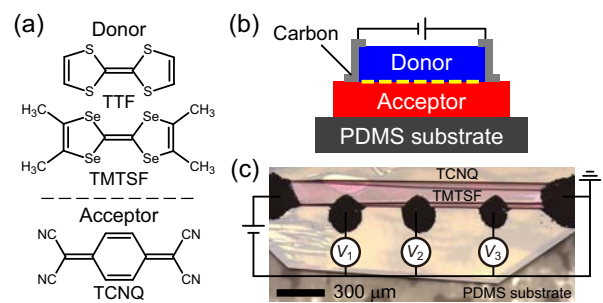

Fig. 1  M. Nakano et al.

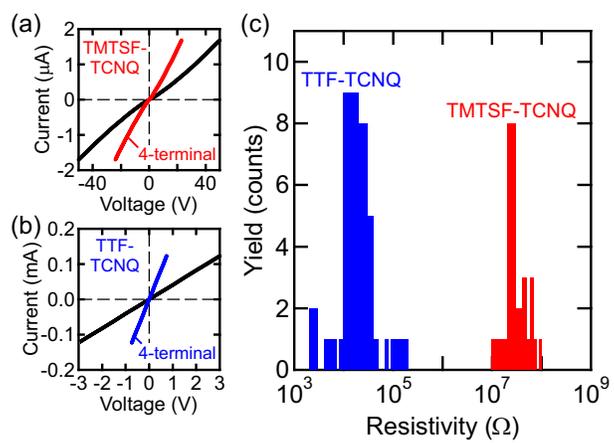

Fig. 2  M. Nakano et al.

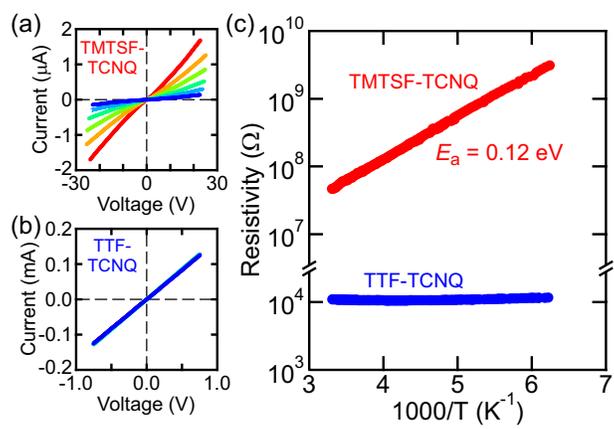

Fig. 3  M. Nakano et al.

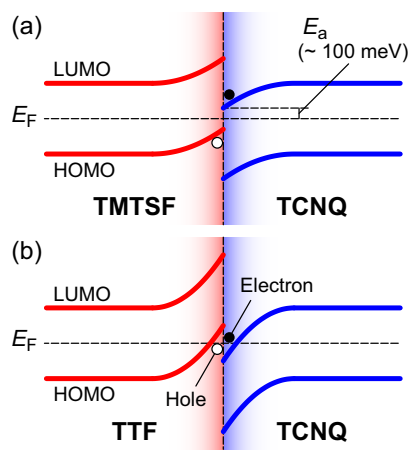

(a)

LUMO

$E_a$
(~ 100 meV)

$E_F$

HOMO

**TMTSF**    **TCNQ**

(b)

LUMO

Electron

$E_F$

HOMO

Hole

**TTF**    **TCNQ**

Fig. 4  M. Nakano et al.